\documentstyle[12pt]{article}
\def \as {\alpha_s}

\def \ms {{\overline{\mbox{MS}}}}

\newcommand{\prepr}[1] {\begin{flushright} {\bf #1} \end{flushright} \vskip 1.5cm}
\newcommand{\titul}[1] {\begin{center}{\large\bf #1 } \end{center}\vskip 1.cm}
\newcommand{\autor}[1] {\begin {center} {\large \lineskip .5em #1 }
                        \end   {center} }
\newcommand{\lugar}[1] {\begin{center} {\it #1} \end{center}}
\newcommand{\abstr}[1] {{\begin{center} \vskip .5cm {\bf Abstract
                        \vspace{0pt}} \end{center}}\begin{quote} #1
                        \end{quote}}
\newcommand{\z}{&&\hspace*{-1cm}}
\begin{document}

\begin{titlepage}
\prepr{US-FT/19-96\\ April 1996}
\titul{The Longitudinal Structure Function F$_L$
\\as a Function of F$_2$ and dF$_2$/dlnQ$^2$
at small x.\\ The Next-to-Leading Analysis}
\autor{A.V. Kotikov\footnote{E-mail:KOTIKOV@LAPPHP8.IN2P3.FR}}
\lugar{Laboratoire de Physique Theorique ENSLAPP\\ LAPP, B.P. 100,
F-74941, Annecy-le-Vieux Cedex, France}
\autor{G. Parente\footnote{E-mail:GONZALO@GAES.USC.ES}}
\lugar{Departamento de F\'\i sica de Part\'\i culas\\
Universidade de Santiago de Compostela\\
15706 Santiago de Compostela, Spain}
\abstr{We present 
a set of formulae to extract the longitudinal
deep inelastic  structure function $F_L$
from the transverse structure function $F_2$ and
its derivative $dF_2/dlnQ^2$ at small $x$. 
Our expressions 
are valid for
any value of $\delta$, being $x^{-\delta}$ the behavior of the parton
densities at low $x$. 
Using $F_2$ HERA data we obtain $F_L$
in the range $10^{-4} \leq x \leq  10^{-2}$ at $Q^2=20$ GeV$^2$.
Some other applications of the formulae are pointed out.
}
\end{titlepage}
\newpage

For experimental studies of hadron-hadron processes on the new, powerful
{\it LHC}
collider, it is necessary to know in detail the
values of the parton (quark and gluon) distributions
(PD) of nucleons, especially at small values of $x$. The basic
information on the quark
structure of nucleons is extracted from the process of deep inelastic
lepton-hadron scattering (DIS). Its differential cross-section has the form:
\begin{eqnarray}
 \frac{d^2 \sigma}{dxdy}~=~ \frac{2 \pi \alpha_{em}^2}{xQ^4}~~ \bigl[
\left( 1 - y + y^2/2
\right) F_2(x,Q^2) - \left(y^2/2\right) F_L(x,Q^2) \bigr],
\nonumber \end{eqnarray}
where $F_2(x,Q^2)$ and $F_L(x,Q^2)$ are the transverse and
longitudinal structure functions (SF), respectively.

The longitudinal SF $F_L(x,Q^2)$ 
is a very sensitive QCD characteristic because it is
equal to zero in the parton model with spin$-1/2$ partons (it is very
large with spin$-0$ partons). 
In addition, at
small values of $x$, $F_L$ data are not yet 
available\footnote{In the time of preparing this article, the H1
  collaboration presented \cite{H1FL} the first (preliminary) measurement
of $F_L$ at small $x$},
as they require a
rather cumbersome procedure (see \cite{1.5}, for example).\\

  In the present article we study the behaviour of
$F_L(x,Q^2)$ at
small values of $x$, using the HERA data \cite{F2H1},
\cite{F2ZEUS} and the method 
\cite{2}
of replacement of the Mellin convolution by ordinary products.
By analogy with the case of the gluon distribution function (see
\cite{KOPA} and its references)
it is possible to obtain the relation between $F_L(x,Q^2)$, $F_2(x,Q^2)$
and $dF(x,Q^2)/dlnQ^2$ at small $x$. Thus, the small $x$ behaviour of
the SF $F_L(x,Q^2)$ can be extracted directly from the measured values
of $F_2(x,Q^2)$ and its derivative without a cumbersome procedure (see
\cite{1.5}).  These extracted values of $F_L$ may be well considered as
{\it new
small $x$ ``experimental data'' of $F_L$}.
When experimental data for $F_L$ at small $x$ become
available with a good accuracy, a violation of the relation will be
an indication
of the
importance of other effects as higher twist contribution and/or
of non-perturbative QCD dynamics at small $x$.

We follow the notation of our previous work in ref. \cite{KOPA}. 
The singlet quark
$s(x,Q^2_0)$ and gluon $g(x,Q^2_0)$ parton distribution functions 
(PDF)\footnote{We use PDF multiplied by $x$
 and neglect the nonsinglet quark distribution at small $x$.}
at some $Q^2_0$ are parameterized by (see, for example, \cite{4}):
\begin{eqnarray} 
p(x,Q^2_0) & = & A_p
x^{-\delta} (1-x)^{\nu_p} (1+\epsilon_p \sqrt{x} + \gamma_p x)
~~~~(p=s,g)
\label{1} 
\end{eqnarray}

The value of $\delta$ is a matter of controversy.
  The ``conventional'' choice is $\delta =0$, which leads to a non-singular
behaviour of the PD
(as for example the $D'_0$ fit in \cite{4})
when $x \rightarrow 0$. Another value,
$ \delta  \sim \frac{1}{2}$, was obtained in the studies performed in
ref. \cite{6} as
the sum of the leading powers of $\ln(1/x)$ in all orders of perturbation
theory (PT) ($D'_-$ fit in \cite{4}). Experimentally, recent NMC data
\cite{7} favor small values of $\delta$. This result is also in
agreement with
present data for ${\bf pp}$ and $\overline {\bf p}{\bf
p}$ total cross-sections (see \cite{8}) and
corresponds to the model of Landshoff and Nachtmann pomeron
 \cite{9} with the exchange of a pair of non-perturbative gluons,
yielding  $\delta =0.086$.  However, the new HERA data
\cite{F2H1, F2ZEUS}  
 prefer $ \delta \geq 0.2$.

From the theoretical side, the type of evolution of the PD
in Eq. (\ref{1}) depends on the value and form of
$\delta ~ (\delta_q = \delta_g)$. For example, a $Q^2$-independent
$\delta$
obeys the
Dokshitzer-Gribov-Lipatov-Altarelli-Parisi (DGLAP)  equation 
when $x^{-\delta} \gg 1$ (see,  for example, \cite{Mar} -
\cite{EKL}). However, if $\delta (Q^2_0) =0$
in some point $Q^2_0 \geq 1 GeV^2$ (see \cite{7}, \cite{BF},
\cite{KOTILOWX}) , the behaviour $p(x,Q^2) \sim Const$
($p=(s,g)$) is not compatible with DGLAP equation and a more singular
behaviour is generated. 

If we restrict the analysis to a Regge-like
form of structure functions, one obtains (see \cite{KOTILOWX})
$$ p(x,Q^2) \sim x^{-\delta_p(Q^2)} $$
with next-to-leading order (NLO) $\delta_q(Q^2) \neq
\delta_g(Q^2)$ intercept trajectories.
 
Without any restriction the double-logarithmical behaviour, i.e.
\begin{eqnarray}
 p(x,Q^2) \sim \exp{\biggl(\frac{1}{2}
   \sqrt{\delta_p(Q^2)ln\frac{1}{x}}\biggr)}  
\label{2} \end{eqnarray}
is generated. 

At NLO and for $f=4$ active quarks one has:
$$\delta_g(Q^2)~=~ \frac{36}{25} t -\frac{91096}{5625} l,~~
\delta_q(Q^2)~=~\delta_g(Q^2)  -20 l$$
where 
$t~=~ln(\alpha(Q^2_0)/\alpha(Q^2))$ and $l~=~
\alpha(Q^2_0) - \alpha(Q^2)$ \footnote{Hereafter
we use $ \alpha(Q^2)= \alpha_s(Q^2)/{4 \pi}$ .}.

As our goal is the extraction of $F_L$ without theoretical
restrictions, we will consider both, the Regge-like behaviour
$ p(x,Q^2) \sim x^{-\delta } $ (if $x^{-\delta} \gg 1$)
and the non-Regge dependence of Eq. (\ref{2}) (if $\delta(Q^2_0)=0$).\\

{\bf 1.}
 Assuming the {\it Regge-like behaviour} for the gluon distribution and
 $F_2(x,Q^2)$ at $x^{-\delta} \gg 1$:
$$g(x,Q^2)  =  x^{-\delta} \tilde g(x,Q^2),~~ 
F_2(x,Q^2)  =  x^{-\delta} \tilde s(x,Q^2), $$
 we obtain the
following equation for the $Q^2$ derivative of
the SF $F_2$:
 \begin{eqnarray} 
\frac{dF_2(x,Q^2)}{dlnQ^2}  &=&
-\frac{1}{2} 
x^{-\delta} \sum_{p=s,g}
\Bigl( 
 r^{1+\delta}_{sp}(\alpha) ~\tilde p(0,Q^2) +
 r^{\delta}_{sp}(\alpha)~ x \tilde p'(0,Q^2)  +
 O(x^{2}) \Bigr) \nonumber \\ 
F_L(x,Q^2)  &=&
x^{-\delta} \sum_{p=s,g}
\Bigl( 
 r^{1+\delta}_{Lp}(\alpha) ~\tilde p(0,Q^2) +
 r^{\delta}_{Lp}(\alpha)~ x \tilde p'(0,Q^2)  +
 O(x^{2}) \Bigr) , 
\label{2.1} \end{eqnarray}
 where $r^{\eta}_{sp}(\alpha) $ and $r^{\eta}_{Lp}(\alpha) $
 are the combinations
of the anomalous dimensions (AD) of Wilson operators 
$\gamma^{\eta}_{sp}= \alpha \gamma^{(0),\eta}_{sp} + \alpha ^2 
\gamma^{(1),\eta}_{sp} + O(\alpha ^3)$
and Wilson 
coefficients\footnote{Because
  we consider here $F_2(x,Q^2)$ but not the singlet quark
  distribution.} $ \alpha B_L^{p,\eta} 
\Bigl(1+ \alpha R_L^{p,\eta} \Bigr)  + O(\alpha ^3)$
and $ \alpha B_2^{p,\eta}  + O(\alpha ^2)$
of the $\eta$
"moment"  (i.e., the corresponding variables extended
from integer values of argument to non-integer ones):
\newpage
 \begin{eqnarray} 
r^{ \eta}_{Ls}(\alpha) ~&=&~ \alpha 
  B^{s,\eta}_{L} \biggl[1 + \alpha \biggl(
  R^{s,\eta}_{L} - B_2^{s,\eta}   \biggr) 
  \biggr] + O(\alpha ^3)  \nonumber  \\  
r^{\eta}_{Lg}(\alpha) ~&=&~ \frac{e}{f}  \alpha 
  B^{g,\eta}_{L} \biggl[1 + \alpha \biggl(
  R^{g,\eta}_{L} - B_2^{g,\eta} B^{s,\eta}_{L} / B^{g,\eta}_{L}   \biggr) 
  \biggr]   + O(\alpha ^3) \nonumber \\ 
r^{ \eta}_{ss}(\alpha) ~&=&~ \alpha 
  \gamma^{(0),\eta}_{ss} + \alpha^2 \biggl(
  \gamma^{(1),\eta}_{ss} + B_2^{g,\eta} \gamma^{(0),\eta}_{gs} +
  2\beta_0 B_2^{s,\eta}
  \biggr) + O(\alpha ^3)  \label{2.2}  \\  
r^{\eta}_{sg}(\alpha) ~&=&~ \frac{e}{f} \biggl[ \alpha 
  \gamma^{(0),\eta}_{sg} + \alpha^2 \biggl(
  \gamma^{(1),\eta}_{sg} + B_2^{s,\eta} \gamma^{(0),\eta}_{sg}
                         + B_2^{g,\eta} \bigl(2\beta_0 + 
  \gamma^{(0),\eta}_{gg} - \gamma^{(0),\eta}_{ss} \bigr) \biggr)
  \biggr] + O(\alpha ^3) \nonumber  \end{eqnarray} 
 and
$$ \tilde p'(0,Q^2) \equiv \frac{d}{dx} \tilde p(x,Q^2) \mbox{ at }
x=0,$$
where $e = \sum_i^f e^2_i$ is
the sum of squares of 
quark charges.

With accuracy of $O(x^{2-\delta})$,  
we have for Eq.(\ref{2.1}) 
 \begin{eqnarray}    \z    \frac{dF_2(x,Q^2)}{dlnQ^2}  = -\frac{1}{2} \biggl[
r^{1+\delta}_{sg} {(\xi_{sg})}^{-\delta}
   g(x/\xi_{sg},Q^2) + r^{1+\delta}_{ss}
F_2(x,Q^2)  + (r^{\delta}_{ss} - r^{1+\delta}_{ss}) x^{1-\delta}
\tilde s'(x,Q^2)
 \biggr] \nonumber \\ \z
~~~~~~~ ~~~~~~~~
+~ O(x^{2-\delta}) \label{5} \\
   \z \nonumber \\ \z   
~~F_L(x,Q^2)  = r^{1+\delta}_{Lg} {(\xi_{Lg})}^{-\delta}
   g(x/\xi_{Lg},Q^2) + r^{1+\delta}_{Ls}
F_2(x,Q^2)  + (r^{\delta}_{Ls} - r^{1+\delta}_{Ls}) x^{1-\delta}
\tilde s'(x,Q^2) \nonumber \\ \z
~~~~~~~ ~~~~~~~~
+~ O(x^{2-\delta}) , \label{5.1}
\end{eqnarray}
with $\xi_{sg} =  r^{1+\delta}_{sg} /r^{\delta}_{sg}$ and 
$\xi_{Lg} =  r^{1+\delta}_{Lg} /r^{\delta}_{Lg}$.

From Eq.(\ref{5}) and (\ref{5.1}) one can obtain $F_L$ as a function
of $F_2$ and the derivative. 

\begin{eqnarray} 
F_L(x, Q^2)  &=& -
\xi ^{\delta}
  \biggl[ 2 \frac{ r^{1+\delta}_{Lg}}{ r^{1+\delta}_{sg}} 
\frac{d F_2(x \xi , Q^2)}{dlnQ^2}
+ \biggl( r^{1+\delta}_{Ls} - \frac{ r^{1+\delta}_{Lg}}{ r^{1+\delta}_{sg}}
r^{1+\delta}_{ss} \biggr) F_2(x \xi ,Q^2)  \nonumber \\ 
 &+&~  
O(x^{2-\delta},\alpha x^{1-\delta}) \biggr], 
\label{8}\end{eqnarray}
where the result is restricted to $O(x^{2-\delta}, \alpha
 x^{1-\delta})$. 

To arrive to the above equation we have 
performed the substitution
$$ \xi_{sg}/\xi_{Lg} \to  \xi =
\gamma_{sg}^{(0),1+\delta} B^{g,\delta}_L /\gamma_{sg}^{(0),\delta}
B^{g,1+\delta}_L  $$ 
and neglected the term $\sim \tilde s'(x \xi_{sg},Q^2)$.

This replacement is very useful. The NLO AD
$\gamma_{sp}^{(1),n}$ are 
singular\footnote{In the case of replacement Mellin convolution
  by ordinary product these singularities transform to logarithmically
  increasing terms (see \cite{KOTIYF93} and \cite{2}).}
 in both points, $n=1$ and $n=0$, and their presence
into the arguments of $\tilde p(x,Q^2)$ makes the numerical
agreement between this approximate formula and the exact calculation worse
(we have
checked this point using some MRS sets of parton
distributions).

Using NLO approximation of $r^{1+\delta}_{sp}$ and $r^{1+\delta}_{Lp}$
  we easily obtain\footnote{The LO analysis was given already in
    \cite{KOTIJETP95}}  the
final results for F$_L(x,Q^2)$:
\begin{eqnarray} \z  F_L(x, Q^2)  = - 2 
\frac{B^{g,1+\delta}_L \Bigl(1+ \alpha \overline R^{g,1+\delta}_L \Bigr)}{ 
\gamma^{(0),1+\delta}_{sg} + 
\overline \gamma^{(1),1+\delta}_{sg} \alpha } \xi ^{\delta}  
\biggl[
\frac{d F_2(x \xi, Q^2)}{dlnQ^2} \nonumber \\ 
\z  ~~ ~~~~~~~~~~ ~+~ \frac{\alpha}{2}  
\biggl( \frac{B^{s,1+\delta}_L}{B^{g,1+\delta}_L} 
\gamma^{(0),1+\delta}_{sg} - \gamma^{(0),1+\delta}_{ss} \biggr) 
F_2(x \xi,Q^2) \biggr]
+
O(\alpha^2,x^{2-\delta},\alpha x^{1-\delta}) 
\label{9.1} \\
\z 
\nonumber \\ 
\z  F_L(x, Q^2)  = - 2 
\frac{B^{g,1+\delta}_L \Bigl(1+ \alpha \overline R^{g,1+\delta}_L \Bigr)}{ 
\gamma^{(0),1+\delta}_{sg} + 
\overline \gamma^{(1),1+\delta}_{sg} \alpha } 
\biggl[
\frac{d F_2(x, Q^2)}{dlnQ^2} \nonumber \\ \z
~~~~~~~~~~~~ ~+~ \frac{\alpha }{2}  
\biggl( \frac{B^{s,1+\delta}_L}{B^{g,1+\delta}_L}
\gamma^{(0),1+\delta}_{sg} - \gamma^{(0),1+\delta}_{ss} \biggr) 
 F_2(x,Q^2) \biggr]
+
O(\alpha^2, x^{1-\delta}), 
\label{9.2} \end{eqnarray}
where
$$\overline \gamma^{(1),\eta}_{sg} ~=~  \gamma^{(1),\eta}_{sg} + 
B_2^{s,\eta} \gamma^{(0),\eta}_{sg}
+
B_2^{g,\eta} \bigl(2\beta_0 + 
  \gamma^{(0),\eta}_{gg} - \gamma^{(0),\eta}_{ss} \bigr),~~
\overline R^{g,\eta}_L ~=~ R^{g,\eta}_L - B^{g,\eta}_2 
\frac{B^{s,\eta}_L}{B^{g,\eta}_L} $$

In principle 
any equation from above formulae (\ref{9.1}), (\ref{9.2}) may be used,
because there is
a strong cancelation between the shifts in the arguments of the function
$F_2$ and its derivative, and the shifts in the coefficients in 
front of them.
The difference lies in the degree of accuracy one can reach with them,
which depends on the $x$ and $Q^2$ region of interest.

For concrete values of $\delta = 0.5$ 
and $\delta = 0.3$ we obtain (for f=4 and $\overline{MS}$ scheme):
\begin{eqnarray} \z  
~\mbox{ if }~ \delta =0.5 \nonumber \\ \z
F_L(x, Q^2)  = \frac{0.87}{ 1 + 22.9 \alpha} \biggl[
  \frac{d F_2(0.70 x , Q^2)}{dlnQ^2} + 
4.17 \alpha F_2(0.70 x,Q^2) \biggr]  +
O(\alpha^2,x^{2-\delta},\alpha x^{1-\delta}) 
\label{10.1} 
%
\\ \z
F_L(x, Q^2)  = \frac{1.04}{ 1 + 22.9 \alpha} \biggl[
  \frac{d F_2(x , Q^2)}{dlnQ^2} + 
4.17 \alpha F_2( x,Q^2) \biggr] +
O(\alpha^2,x^{1-\delta}) 
\label{10.2} 
\\ \z \nonumber 
\\ \z
~\mbox{ if }~ \delta =0.3 \nonumber \\ \z
F_L(x, Q^2)  = \frac{0.84}{1 + 59.3 \alpha }  \biggl[
 \frac{d F_2(0.48 x, Q^2)}{dlnQ^2} + 3.59 \alpha 
F_2(0.48 x , Q^2) \biggr]
+ O(\alpha^2 \!\!, x^{2-\delta} \!\!,\alpha x^{1-\delta}) 
\!         
\label{10.3}
\\ \z
F_L(x, Q^2)  = \frac{1.05}{1 + 59.3 \alpha } \biggl[
 \frac{d F_2(x, Q^2)}{dlnQ^2} + 3.59 \alpha 
F_2(x , Q^2)  \biggr]
+ O(\alpha^2 \!\!, x^{1-\delta}) 
\!         
\label{10.4} 
\end{eqnarray}\\

{\bf 2.} 
 Assuming the {\it non-Regge-like behaviour} for the gluon distribution and
 $F_2(x,Q^2)$:
$$g(x,Q^2)  =  \frac{
\exp{(\frac{1}{2} \sqrt{\delta_g(Q^2)ln\frac{1}{x}})}  }{
{(2\pi \delta_g(Q^2)ln\frac{1}{x})}^{1/4}       }
 \tilde g(x,Q^2),~~ 
F_2(x,Q^2)  =  \frac{
\exp{(\frac{1}{2} \sqrt{\delta_s(Q^2)ln\frac{1}{x}})}  }{
{(2\pi \delta_s(Q^2)ln\frac{1}{x})}^{1/4}       }
 \tilde s(x,Q^2), $$
 we obtain the
following equation for the $Q^2$ derivative of
the SF $F_2$\footnote{Using a lower approximation $O(x)$ is not very exact,
because in this case $F_2$ and the gluon distribution can contain an
additional factor in the form of a serie $1+ \sum_k
(1/\delta_p/ln(1/x))^k$,
which is determined by boundary conditions (see discussion in
Ref.\cite{BF}). We will not consider the appearance of this factor in
our analysis}:
 \begin{eqnarray}\z \frac{dF_2(x,Q^2)}{dlnQ^2}  =
-\frac{1}{2} 
 \sum_{p=s,g} \frac{
\exp{(\frac{1}{2} \sqrt{\delta_p(Q^2)ln\frac{1}{x}})}  }{
{(2\pi \delta_p(Q^2)ln\frac{1}{x})}^{1/4}       }
\Bigl(  \tilde
 r^{1}_{sp}(\alpha) ~\tilde p(0,Q^2) +
 O(x^{1}) \Bigr) ,\label{13} \\
\z
~~F_L(x,Q^2)  =
 \sum_{p=s,g} \frac{
\exp{(\frac{1}{2} \sqrt{\delta_p(Q^2)ln\frac{1}{x}})}  }{
{(2\pi \delta_p(Q^2)ln\frac{1}{x})}^{1/4}       }
\Bigl(  \tilde
 r^{1}_{Lp}(\alpha) ~\tilde p(0,Q^2) +
 O(x^{1}) \Bigr) ,\label{13.1}
\end{eqnarray}
 where $ \tilde r^{1 }_{sp}(\alpha) $ and $ \tilde r^{1 }_{Lp}(\alpha)
 $ can be obtained from
 corresponding functions   
$ r^{1+\delta }_{sp}(\alpha) $ and $ r^{1+\delta }_{sp}(\alpha) $,
respectively, 
 replacing the singular term $1/\delta$
at $\delta \to 0$ by $1/\tilde \delta$:
\begin{eqnarray}
\frac{1}{\delta} \stackrel{\delta \to 0}{\to} \frac{1}{\tilde \delta}
~=~ \sqrt{\frac{ln(1/x)}{\delta_p(Q^2)}} 
- \frac{1}{4\delta_p(Q^2)} \left[ 1 + \sum_{m=1}^{\infty}
\frac{1 \times 3 \times ... \times
(2m-1)}{\left(4\sqrt{\delta_p(Q^2) \ln(1/x)}\right)^m} \right]
\label{14} \end{eqnarray}
 The singular term appears only in the NLO part of the AD
$\gamma^{(1),1+\delta}_{sp}$ and the longitudinal Wilson coefficients
$R^{p,1+\delta}_{L}$
in Eq. (\ref{2.2}).
The replacement (\ref{14}) corresponds to the following transformation:
\begin{eqnarray} 
\gamma^{(1),1+\delta}_{sp} &\equiv & \hat 
\gamma^{(1),1}_{sp} \frac{1}{ \delta} +\breve \gamma^{(1),1+\delta}_{sp}
~~ \stackrel{\delta \to 0}{\to} ~~  \tilde \gamma^{(1),1}_{sp} = \hat 
\gamma^{(1),1}_{sp} \frac{1}{\tilde \delta} +\breve \gamma^{(1),1}_{sp}
\nonumber  \\ 
R^{p,1+\delta}_{L} &\equiv & \hat 
R^{p,1}_{L} \frac{1}{ \delta} +\breve R^{p,1+\delta}_{L}
~~~~ \stackrel{\delta \to 0}{\to} ~~  \tilde R^{p,1}_{L} = \hat 
R^{p,1}_{L} \frac{1}{\tilde \delta} +\breve R^{p,1}_{L}
 \label{15} \end{eqnarray}
where $\hat \gamma^{(1),1}_{sp}$ ($\hat R^{p,1}_{L}$) and 
$\breve \gamma^{(1),1+\delta}_{sp}$ ($\breve R^{p,1+\delta}_{L}$)
are the coefficients corresponding to singular and regular parts of
$\gamma^{(1),1+\delta}_{sp}$ ($R^{p,1+\delta}_{L}$), respectively.

We restrict here our calculations to $O(x)$ because at
$O(x^2)$ one obtains an additional factor
in front of the function $F_2$ and its derivative, which
complicates very much the final formulae.

Repeating the analysis of the
previous section step by step using the replacement (\ref{15}), we get
(for f=4):
\begin{eqnarray} \z  
  F_L(x, Q^2)  = 
\frac{1}{ (1 + 30 \alpha [
1/\tilde \delta - \frac{116}{45}])  } \biggl[
\frac{d F_2(x, Q^2)}{dlnQ^2} 
+ \frac{8}{3} \alpha F_2(x, Q^2)   \biggr] ~+~ 
O(\alpha^2, x) 
\label{11.1} 
\end{eqnarray} \\

We have combined
equations (\ref{9.2}) and (\ref{11.1}) in a more general
formula valid for any value of $\delta$:
\begin{eqnarray} 
\z  F_L(x, Q^2)  = - 2 
\frac{B^{g,1+\delta}_L \Bigl(1+ \alpha \tilde R^{g,1+\delta}_L \Bigr)}{ 
\gamma^{(0),1+\delta}_{sg} + 
\tilde \gamma^{(1),1+\delta}_{sg} \alpha } 
\biggl[
\frac{d F_2(x, Q^2)}{dlnQ^2} \nonumber \\ \z
~~~~~~~~~~~~ ~+~ \frac{\alpha }{2}  
\biggl( \frac{B^{s,1+\delta}_L}{B^{g,1+\delta}_L}
\gamma^{(0),1+\delta}_{sg} - \gamma^{(0),1+\delta}_{ss} \biggr) 
 F_2(x,Q^2) \biggr]
+
O(\alpha^2, x^{1-\delta}), 
\label{13.1} 
\end{eqnarray}
where
$\tilde \gamma^{(1),1+ \delta}_{sg}$ and $\tilde R^{g,1+ \delta}_{L}$
 coincide with $\overline \gamma^{(1),1+\delta}_{sg}$ and 
$\overline R^{g,1+\delta}_{L}$, respectively,
  with the replacement:
\begin{eqnarray}
\frac{1}{\delta} \to \int^1_x \frac{dy}{y}~ \frac{g(y,Q^2)}{g(x,Q^2)} 
\label{14.1} 
\end{eqnarray}

In the cases $x^{-\delta} \gg Const$ and $\delta \to 0$, the r.h.s. of 
(\ref{14.1}) leads to $1/\delta $ and $1/\tilde \delta $,
respectively.\\

{\bf 3.}
In Fig. 1 it is shown the accuracy of Eqs. (\ref{10.1})-
(\ref{10.4}) 
and (\ref{11.1})
in the reconstruction of $F_L$ at
various $\delta $ values from MRS sets
at Q$^2$=20 GeV$^2$.
We have chosen for this test MRS(D$_{0}$) ($\delta$=0), MRS(D$_{-}$) 
($\delta$=0.5) and MRS(G) ($\delta$=0.3) as three representative
densities (see ref. \cite{4}
and references therein).
It can be observed in Fig. 1a that using the formulae (\ref{10.1})
and (\ref{10.2})
one gets very good
agreement with the input parameterization MRS(D$_{-}$) (less than 1
$\%$) 
at low x. 

Fig. 1b  shows the degree of accuracy of the reconstruction formulae
(\ref{10.3}) and (\ref{10.4}) with $\delta=0.3$. Here one should expect
the set MRS (G) to give
also a very good ($\sim 1 \%$ level)
agreement, 
however this is not the case because set (G)
distinguishes the exponents of the sea-quark part $\delta_s \sim 0$
from the gluon density ($\delta_g=0.3$). Thus, Eq. (12) might be slightly
modified to treat this case. Note that the agreement is improved
when $x$ values decrease because the relative
importance of the quark contribution becomes smaller.

Fig. 1c deal with the case $\delta=0$. As in Fig. 1a, one
can observe a very 
good accuracy 
in the reconstruction when $Q_0^2$ is closed to
that of the test parameterization (4 $GeV^2$ for MRS set).
Notice also the lost of accuracy at high $x$ due
to the importance of the O($x$) terms neglected in Eq.(\ref{11.1}).
 
With the help of Eqs. (\ref{10.1}), (\ref{10.3}) and (\ref{11.1})
we have extracted the longitudinal SF $F_L(x,Q^2)$
from HERA data, using the slopes dF$_2$/dlnQ$^2$
determined in ref. \cite{H1PREP95} and ref. \cite{ZEUSGLU95}. 
When H1 data are used, the value of F$_2$ in Eq. (\ref{10.1}) was
directly taken from the parameterization given by H1 in ref. \cite{F2H1}.
With ZEUS data we substitute directly the F$_2$ values
presented in table 1 of ref. \cite{ZEUSGLU95}.
We have checked that the
use of the H1 parameterization for $F_2$ when dealing with ZEUS data,
does not change significantly the $F_L(x,Q^2)$ result.
Another input ingredient in the extraction formulae
is $\alpha_s(Q^2)$. We have used the NLO QCD approximation with
$\Lambda_{\ms} = 225$ MeV, even though the
results are no very sensitive to this value. For example, a variation
in $\Lambda$ of around $\pm 50$ MeV changes the results less that a 1 $\%$.

Figure 2 shows the extracted values of the longitudinal SF 
and the prediction in QCD using
MRS sets (G), (D$_-$) and (D$_0$) and the O($\alpha_s^2$)
coefficients calculated in ref. \cite{FLNLO}.
For comparison we have included in the same figure the results from
different formulae.
In Fig. 2a the points extracted with
$\delta=0$ and $\delta=0.5$ are spread over a band which could be
considered as an indication of the theoretical uncertainty of the method,
if $\delta$ were completely unknown. However, in
a realistic situation, the uncertainty should be smaller
if one could restrict in advance the value of $\delta$,
as it is discussed below.

On the other hand, the deviation between the data points using $\delta=0$
and the prediction of $F_L$  from MRS(D$_0$) parton
distributions, is a signal that the formula is inadequate
for the extraction of $F_L$. In this case the origin of the discrepancy
is not
clear. It could be due to the importance of other contributions,
not considered in the formula, or perhaps simply that $\delta$
is large.  

In general it can be observed that the agreement, within the errors, 
with the calculation from sets MRS(G) and MRS(D$_-$)
is excellent. 
There is also a relative good agreement with a preliminary
experimental H1 point for $F_L$ ref. \cite{H1FL}, if one takes into
account the
systematic error, not shown in Fig. 2a.\\

{\bf 4.}
In summary, we have presented  Eqs. (\ref{9.1})-(\ref{10.3}) for the
extraction of the longitudinal SF $F_L$ at small $x$ from the SF
$F_2$ and its
$Q^2$ derivative. These equations provide the possibility of the non-direct
determination of $F_L$. This is important since the direct
extraction of $F_L$ from experimental data is a cumbersome procedure
(see \cite{1.5}).
Moreover, the fulfillment of  Eqs. (\ref{9.1})-(\ref{10.4}) in DIS
experimental data is a cross-check of perturbative QCD at small values of
$x$.

We have found, as in the case of the gluon extraction formulas \cite{KOPA}, 
that for singular type of partonic densities
the results do not depend practically on the concrete value
of the slope $\delta $, due to a cancelation of that dependence
between certain coefficients.
However, when
$\delta \to 0$, the coefficients in front of
$dF_2(x,Q^2)/dlnQ^2$ and $F_2(x,Q^2)$ have singularities
leading to terms $\sim \sqrt{ln(1/x)}$\footnote{This happens
in the framework of
the double-logarithmic asymptotic. The singularities lead to terms
$\sim ln(1/x)$ in the case of Regge-like asymptotic (see
\cite{KOTILOWX}, \cite{KOTIYF93}, \cite{2}).}.
In this case there is a strong correlation between the results and the 
concrete form of small $x$ asymptotics of $F_2(x,Q^2)$.

Consequently,
before to apply these formulae, some
study of the
experimental data is necessary in order to verify the type of $F_2(x,Q^2)$
behavior at $x \to 0$ (i.e. the value of $\delta $).
Note that this study should be
done at a fixed value of $Q^2$ and it does not require
the knowledge of the quark and gluon content.
For example, in Ref. \cite{EKL} it
was suggested the determination of the slope from the observable
$dF_2(x,Q^2)/dlnx$, which was measured in \cite{H1PREP95}.

\vskip 0.5 cm
%
%
%
%
\noindent{\bf Acknowledgments }

This work was supported in part by CICYT and by Xunta de Galicia.
We are grateful to J.W. Stirling for providing
the parton distributions used in this work, and to 
A. Bodek and M. Klein for discussions.

%
%

%
%
%
\noindent{\bf Figure captions }
\vspace{1.cm}

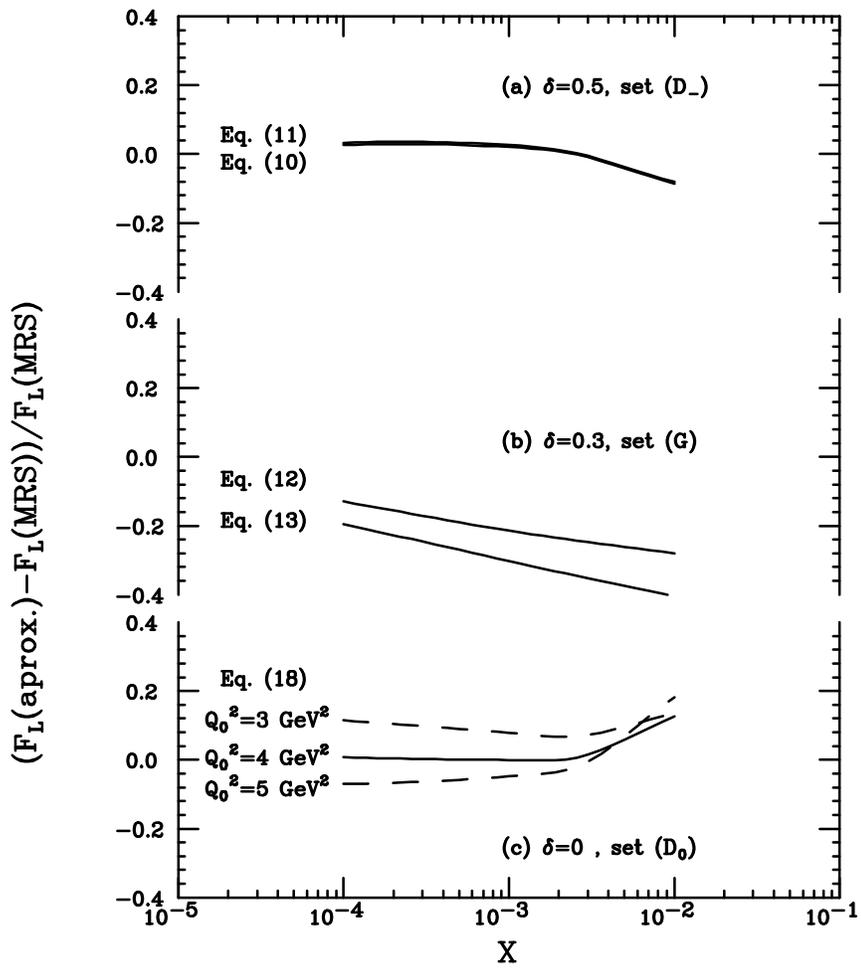
\noindent{Figure 1:
Relative difference between the reconstructed longitudinal SF
using formulae in text and different input parameterizations
at $Q^2$= 20 GeV$^2$.}

\vspace{1.cm}

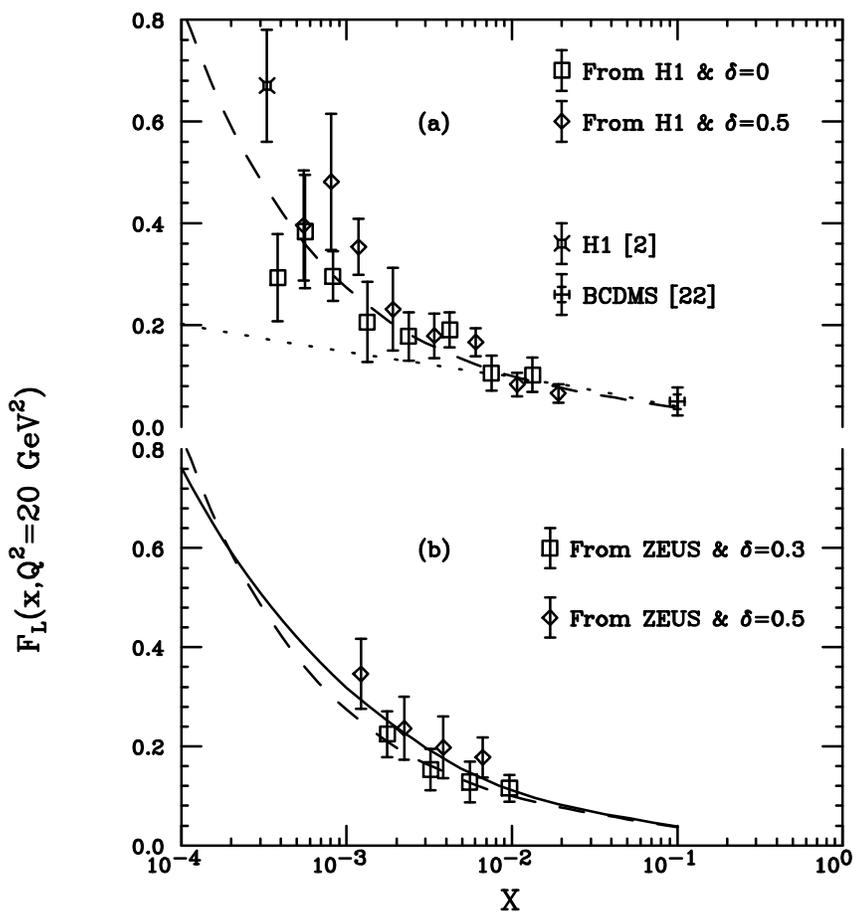
\noindent{Figure 2: The longitudinal SF $F_L$. The points were extracted
from Eqs. (\ref{10.1}) and (\ref{11.1})  
using H1 \cite{H1PREP95} data (Fig. 2a) and from Eqs. (\ref{10.1})
and (\ref{10.3}) using ZEUS \cite{ZEUSGLU95} data (Fig. 2b).
Solid, dashed and dotted lines are the calculation from sets MRS(G),
MRS(D$_-$) and MRS(D$_0$) \cite{4}
using O($\alpha_s^2$) corrections. It is also shown a BCDMS data point at
$x=0.1$ and a preliminary H1 data point.}

\vspace{1.cm}
\end{document}